# Hidformer: Transformer-Style Neural Network in Stock Price Forecasting

Kamil Ł. Szydłowski[1][0009−0003−8398−5115] and
Jarosław A. Chudziak[1][0000−0003−4534−8652]

Institute of Computer Science, Warsaw University of Technology, Poland
{kamil.szydlowski.stud,jaroslaw.chudziak}@pw.edu.pl

**Abstract.** This paper investigates the application of Transformer-based neural networks to stock price forecasting, with a special focus on the intersection of machine learning techniques and financial market analysis. The evolution of Transformer models, from their inception to their adaptation for time series analysis in financial contexts, is reviewed and discussed. Central to our study is the exploration of the Hidformer model, which is currently recognized for its promising performance in time series prediction. The primary aim of this paper is to determine whether Hidformer will also prove itself in the task of stock price prediction. This slightly modified model serves as the framework for our experiments, integrating the principles of technical analysis with advanced machine learning concepts to enhance stock price prediction accuracy. We conduct an evaluation of the Hidformer model's performance, using a set of criteria to determine its efficacy. Our findings offer additional insights into the practical application of Transformer architectures in financial time series forecasting, highlighting their potential to improve algorithmic trading strategies, including human decision making.

## 1 Introduction

The stock market, commonly perceived as a venue for investment and potential profit (or loss) by the average person, is underpinned by the intricate field of market analysis. This discipline is dedicated to forecasting financial market behaviors using diverse methodologies. Some analysts rely on economic fundamentals to evaluate companies, while others interpret market trends directly from charts, suggesting that these visuals encapsulate all necessary information [10].

It should be noted that the topic of stock market price prediction has consistently maintained its popularity in the field of artificial intelligence research over years. The most popular proposals in the literature in recent years are artificial neural networks - especially models and gates based on the principle of recursion, but also others: both traditional ones (including convolutional networks) and highly specialized architectures, which with the help of knowledge transfer can achieve high results in this field as well [3, 5, 11]. So far, analyses have shown that hybrid projects have the potential to achieve better results than mono solutions [5, 11]. Currently, however, complex models such as the Transformer are also being studied for this purpose [13].

This article aims to evaluate *Hidformer's* [8] effectiveness in stock price forecasting as a modification of the Transformer model [12] for time series prediction. Our experiments demonstrate that *Hidformer* either outperforms or matches existing basic models based on standard metrics and visual analytics. In the first place we will briefly outline the concept of *technical analysis* of stock markets (Section 2), and then we will provide an overview of Transformer neural networks, moving from the original model to ideas focused on time series processing, including stock market prediction (Section 3). In particular, we will describe the *Hidformer* model (Section 4), currently considered as the very promising NN architecture in time series prediction, which will serve as the basis for experiments (Section 5). And finally, we will discuss results of the conducted experiments, outlining the scope for future research as well (Section 6).

## 2  Analysis of Financial Stock Markets

Market dynamics analysis aims to forecast the behavior of financial markets [10]. This section recalls the main notions and elements of technical analysis, as well as the assumptions of basic deep learning techniques in this field.

### 2.1  Preliminaries

Technical analysis studies market behaviors, primarily using charts, to predict future price trends. This approach relies on the interpretation of price and volume data. By examining historical price movements and trading volumes, analysts identify patterns and trends that suggest future market behavior.

The justification for using technical analysis as a tool for market analysts is based on three premises [10]. First, **the market discounts everything**, meaning that all factors affecting market prices are reflected in those prices. Second, **prices move in trends**, indicating that an existing trend will continue until it reverses. And third, **history repeats itself**, based on the assumption that historical patterns, such as the head and shoulders formation [10], are believed to reoccur and can predict future market behavior. However, we are aware of the computational and economic limitations of these assumptions [1].

Technical analysis can be applied using various valuation frequencies, including high-frequency valuations, which analyze market movements in fractions of a second; inter-day valuations, which consider closing prices over multiple days; and intra-day valuations, which examine price movements within a single trading day. For our study, we assume a set of parameters based on daily valuations, which provide a balance between capturing market trends and managing data complexity [11].

Daily valuations could be based on six components (for a very basic approach see Fig. 1): Opening and Closing Price, the valuation of a given instrument at the first and last quotation of the day; Highest and Lowest Price, the highest and lowest price of a given instrument during the day; Adjusted Closing Price, the true value of a given instrument calculated after distributing dividends; and Volume, the number of contracts concluded.

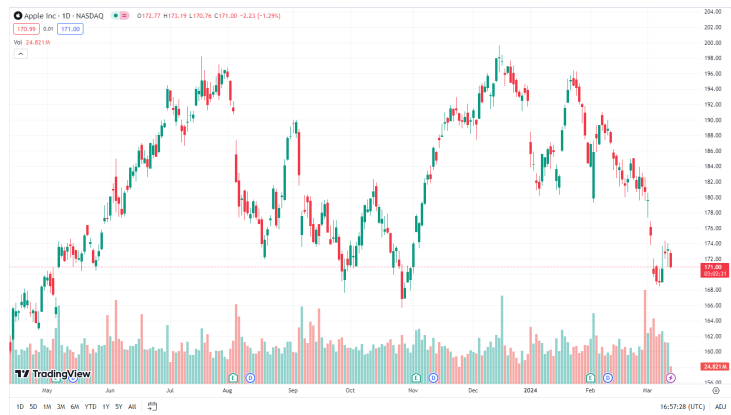

**Fig. 1.** Sample stock chart (https://www.tradingview.com/).

The aforementioned daily stock quote corresponds, on a logical level, to the format of the data we use. In our study, we will show that our model will learn the above premises of the justification for using technical analysis *tabula rasa*.

### 2.2 Market analysis with deep learning neural networks

In recent years, with the intensive development of deep learning methods and algorithms, many solutions have been proposed for stock price prediction and algorithmic trading. Among the most widely used and discussed were [2, 5]:

1. **Convolutional Neural Networks (CNN)**. In the context of stock prices, they can be used as analyzers of sequential data [5, 6]. In other words, if there are distinctive raw features in the data, the network is capable of extracting standard formations that can be the basis of a strategy.
2. **Recurrent Neural Networks (RNN)**. They process individual input data elements and their connections with previous elements. In this way, they retain information about the context [4, 6]. As a result, they are potentially a good choice for predicting stock data, as changes in quotes over a given time are often related to previous trends [5, 6].
3. **Long Short-Term Memory (LSTM)**. Given that stock price prediction might be represented as a nonlinear process, it seems that the LSTM gate, which is one of the specialized recurrent architectures, is capable of effectively uncovering correlations between nonlinear time series and the prediction target [5].
4. **Deep Neural Networks (DNN)**. Discerning their proficiency in handling nonlinear problems, it must be emphasized that these models might also be used for stock price prediction [5].
5. **Transformers**: Recently, transformers have emerged as a powerful architecture for stock price prediction due to their ability to handle sequential data efficiently and capture long-term dependencies [13].

## 3 Related work

After the innovative Transformer neural network model [12] was presented in 2017, it was acknowledged that this model represented a significant advancement in the use of the attention mechanism. Previously used recurrent architectures exhibited major limitations when working with long sequences. The Transformer architecture solved these issues by introducing self-attention, multi-head attention, and positional encoding, which improved the model's ability to encode and process sequential data efficiently.

Among the applications explored by the authors were machine translation and syntactic analysis of natural language. However, they also recognized that this model was worth testing in multimodal sequences (images, sound, and video). The key to the Transformer's success turns out to be its ability to process sequential data, which is of particular significance in our considerations.

### 3.1 Transformer model in time series prediction

The initial effort to adapt the Transformer model for time series analysis occurred in 2019 [7]. Despite identifying the primary limitations of the architecture, this pioneering work also proposed solutions to address these challenges:

1. **Locality-agnostics**. The point-wise dot-product self-attention turned out to be insensitive to the local context of a given point and was also very prone to single anomalies - outliers as a result of noise. These issues were resolved by the mechanism of convolutional self-attention, which considered several local values.
2. **Memory bottleneck**. Due to the fact that memory complexity grows quadratically, it became apparent that there was a lack of capability to model long, much longer than sentences, sequences. However, even in this context, a correction was introduced by replacing full self-attention with sparse self-attention, which only considered the most valuable elements of the sequence.

Independently of the aforementioned studies, in 2021, an attempt was made to adapt the original concept of the Transformer for time series prediction. In particular, the authors decided to [9]:

1. **Omit the embedding layer**. Since we deal with numerical data, they already constitute embeddings themselves, unlike words.
2. **Add a time encoding layer**. This layer models the temporal relationships between elements in the sequence better than positional encoding alone.
3. **Fix training-related issues**. The last *softmax* layer was removed from the decoder and MSE (Mean Squared Error) was used as the loss function due to working with data from a continuous distribution (numbers), which is not a discrete one (words).

In the same year (2021) the first serious Transformer-type architecture for processing time series was described. It was named *Informer* [16] (see Fig. 2).

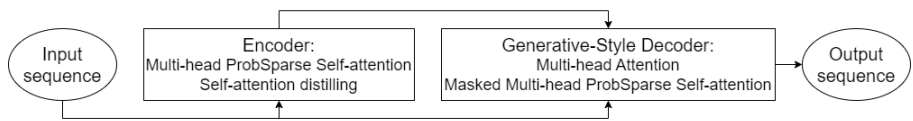

**Fig. 2.** Informer model - conceptual overwiew.

The leading improvements were:

1. **A ProbSparse self-attention mechanism** which reduces the time and memory complexity of the algorithm to $O(L \log L)$, considering $L$-length inputs/outputs.
2. **The self-attention distilling** which is a modification of self-attention that highlights dominating attention by halving cascading layer input, which in turn ensures efficient processing of long input sequences.
3. **The generative-style decoder** whose prediction of long time series occurs in one operation, significantly increasing the inference speed for long time windows despite its conceptual simplicity.

### 3.2 Transformer in predicting stock prices

In 2022, an attempt was made to apply the *Informer* model in the task of predicting stock prices [13]. The authors highlighted the *Informer*'s advantages of the multi-head attention mechanism. Its effectiveness was particularly evident in efficiently capturing relevant information and filtering out irrelevant noise, which are especially abundant in financial time series, and in the strong ability to extract key features, leading to better forecasting performance.

In the research, its performance was compared with traditional deep learning models (CNN, RNN, and LSTM), where *Informer* demonstrated higher accuracy significantly in predicting the first future quote in all back-testing experiments on the main stock market indices worldwide, including CSI 300, S&P 500, Hang Seng Index, and Nikkei 225, what enabled investors to gain excess earnings.

## 4 Hidformer solution for stock price prediction

Below, we present an idea on how to use the *Hidformer* artificial neural network to predict future stock prices based solely on their past quotations. Expanding the experiments presented previous on the Weather, Traffic, and Electricity datasets, we propose a solution similar to what was done with the *Informer* in finance, but this time we enable the model to predict more than one (more than the only next) quotation of an instrument, which is particularly important in the attempt to determine the trend in human decision making.

### 4.1 Stock prices - data and feature engineering

In finance, stocks are categorized into several industries [14]. As a historical price dataset we use few-year price movements of stocks from the only one particular sector since, according to our experience, stock prices in different industries are characterized by different formations, which could even lead to an attempt to induce contradictory patterns when using machine learning. These observed data for each index is a six-dimensional time series of daily open, high, low, close, adjusted close prices and volumes.

We normalize the dataset as follows:

$$\hat{x}_t = \frac{x - x_{min}}{x_{max} - x_{min}} \quad (1)$$

where $\hat{x}_t$ is the normalized price at time $t$ and $x_{max}$ ($x_{min}$) is the maximum (the minimum) value in the entire sample regarding all dimensions of the prices. With this approach, the values in the sample are in the range $[0, 1]$, taking into account the dependencies between the components of the daily valuation of the security and the volume separately.

In each prediction, we use the previous $T_x$ prices and volumes to predict the close price on the next $T_y$ trading days, utilizing a moving window approach.

### 4.2 Proposed model

As mentioned above, we adopt the *Hidformer* model [8], for the task of stock market prediction. With respect to the original contribution, the architecture is as in the canonical Transformer besides the following improvements [8]:

1. **Two-Tower architecture**. The input sequence is processed through two encoders, one operating in the time domain and the other working from the frequency domain perspective.
2. **Segment-and-merge approach**. This segmentation involves creating subsequences from the original series with potential overlap or a step. These subsequences are merged after each encoder block, allowing for better capture of local features.
3. **Recursive and linear attention**. The classic multi-head attention is replaced with a recursive version in the time encoder and a linear model in the frequency encoder.

4. **MLP-Type decoder**. The long time series prediction occurs in one operation as introduced earlier [16].

Thus, the *Hidformer* model operates as follows (see Fig. 3). First, the token segmentation layer generates $N_T$ tokens. Then all six-dimensional tokens are flatten to $N_E$-dimensional embeddings which are processed by the two towers. Each of them consists of $N_B$ blocks. After merging, the output of each block is the input of the next block. However, at the output of the tower there are concatenated outputs of each block. The outputs of the time and frequency tower are concatenated once again and fed into the decoder, which is actually a multi-layer perceptron with $N_D$ layers.

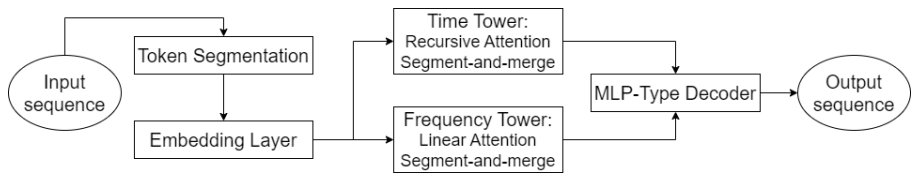

**Fig. 3.** Hidformer model overwiew. [16]

Considering the description above, we believe that after adding improvements related to data processing from the frequency domain perspective, the Transformer-type neural network will prove effective in predicting stock market quotations, as discussed below.

## 5 Experiments

In this section, we conduct experiments with the aim of answering whether the *Hidformer* model is capable of predicting future stock index values, which we verify by comparing its output with the outcome of classic artificial neural networks (CNN, RNN, LSTM and DNN), and confirm through a Mann-Whitney $U$ test. Additionally, we present a visual analysis.

### 5.1 Dataset selection

As a historical price dataset we selected the 43-year price movements from 12/12/1980 to 29/12/2023 of 6 stock indices (see Table 1) which are part of the *Consumer Goods* sector. We used Yahoo Finance as the data source. The training set included the first 95% data (from 12/12/1980 to 31/12/2021), which was used to train model parameters. The last 5% data (from 31/12/2021 to 31/12/2023) as the validation set was used to evaluate the performance of the models.

Table 1. Target stocks.

| Stock symbol | Company |
|---|---|
| AAPL | Apple Inc. |
| KO | The Coca-Cola Company |
| MO | Altria Group, Inc. |
| PEP | Pepsico, Inc. |
| PG | The Proctor & Gamble Company |
| TM | Toyota Motor Corporation |

### 5.2 Hyper-parameters setting

In each prediction, we use the previous $T_x = 128$ prices and volumes to predict the close price on the next $T_y = 128$ trading days. The batch size for the mini-batch training is set as 64. The Adam optimizer with a learning rate of 0.0001 is used for training models. The number of epochs was set as 100 to guarantee the convergence of the training process. To validate, we selected the model chosen as the best during training. The loss function was weighted MSE loss (weights from $T_y = 128$ to 1).

In the *Hidformer* model we set $N_T = 4$ tokens which are flatten to $N_E = 1$-dimensional embeddings. Each tower consists of $N_B = 3$ blocks. The decoder contains $N_D = 2$ layers.

### 5.3 Evaluation criteria

Below we propose criteria to evaluate the performance of the models from two perspectives, having regard to the fact that the effectiveness of a stock price predictor should be measured not only by the difference between the predicted values and the true data, but also by its usefulness in trading [13].

**Prediction accuracy** For prediction accuracy, we compare the network output with the ground truth in the validation set and calculate the three common indicators: Mean Absolute Error (MAE), Mean Squared Error (MSE) and Mean Absolute Percentage Error (MAPE). The lower values of them, the performance is better.

**Net value** In order to evaluate usefulness of the model in trading, we adopt a simple strategy known in the literature [13]. If the predicted close price $\hat{y}_{t+1}$ is larger than the latest observed close price $y_t$, we long one position index (buy one unit of the index); otherwise, we short one position index (sell one unit of the index). Thus, the return at time $t + 1$ is expressed as follows ($y_{t+1}$ is the true next close price):

$$R_{t+1} = \ln \frac{y_{t+1}}{y_t} \times sign(\hat{y}_{t+1} - y_t) \tag{2}$$

So, the net value, representing the total return of the strategy, might be calculated for $t \geq 2$ as follows:

$$NV_t = 1 + \sum_{i=2}^{t} R_i \qquad (3)$$

### 5.4 Key results and analysis

Considering the uncertainty of deep learning methods, we implement 5 independent runs of learning and validation. Based on this, we present means and standard errors of the aforementioned indicators.

**Prediction accuracy** The reported averaged results for all listings from the validation set are presented in Table 2 (the lower the value, the better).

**Table 2.** The means and standard errors of three indicators across 5 independent experiments.

| Model | MAE | MSE | MAPE |
|---|---|---|---|
| CNN | 0.160 (0.001) | 0.039 (0.001) | 67 (1) |
| RNN | 0.198 (0.002) | 0.055 (0.001) | 87 (2) |
| LSTM | 0.189 (0.005) | 0.052 (0.002) | 83 (3) |
| DNN | 0.163 (0.009) | 0.042 (0.004) | 66 (1) |
| **Hidformer** | **0.159 (0.001)** | **0.040 (0.001)** | **66 (1)** |

The preliminary results of the experiments demonstrate that *Hidformer* outperforms in many cases other classic methods significantly (*P*-values lower than 0.01) or achieves at least similar results both from the perspective of absolute errors (MAE and MSE) and relative error (MAPE).

**Net value** Below we present the results of the given trading strategy separately for each index both from the perspective of the average 1-day return (Table 3) and the 2-year trading backtest (Table 4). The higher the value, the better.

**Table 3.** The average return of the trading strategy across 5 independent experiments.

| Model | Net Value | | | | | |
|---|---|---|---|---|---|---|
| | AAPL | KO | MO | PEP | PG | TM |
| CNN | 0.007 | −0.031 | −0.021 | −0.007 | −0.009 | 0.002 |
| RNN | 0.002 | −0.052 | −0.012 | −0.016 | −0.006 | −0.042 |
| LSTM | 0.004 | −0.047 | −0.005 | −0.010 | −0.002 | −0.025 |
| DNN | 0.001 | −0.040 | −0.017 | −0.003 | −0.006 | −0.02 |
| **Hidformer** | **0.007** | **-0.018** | **-0.015** | **0.004** | **0.001** | **0.000** |

The analysis conducts indicate that *Hidformer* achieved a competitive advantage over other methods, as evidenced by often significantly ($P$-values lower than 0.01) higher or similar average returns. Moreover, the analysis shows that also from the perspective of a 2-year backtest, *Hidformer* achieves significantly ($P$-values lower than 0.01) better or the same results compared to other methods.

**Table 4.** The performance of the trading strategy (the means and standard errors) across 5 independent experiments.

| Model | Net Value | | | | | |
|---|---|---|---|---|---|---|
| | AAPL | KO | MO | PEP | PG | TM |
| CNN | 3.4 (3.1) | −10.6 (11.3) | −6.6 (2.7) | −1.7 (3.3) | −2.4 (2.6) | 1.9 (7.6) |
| RNN | 1.6 (0.5) | −18.3 (1.2) | −3.6 (2.9) | −4.9 (0.5) | −1.1 (1.2) | −14.6 (2.0) |
| LSTM | 2.5 (0.3) | −16.6 (0.2) | −0.7 (2.9) | −2.7 (0.2) | 0.3 (0.3) | −14.4 (0.3) |
| DNN | 1.2 (1.7) | −13.8 (8.3) | −5.3 (3.9) | −0.2 (4.9) | −1.2 (5.2) | −6.5 (6.0) |
| **Hidformer** | **3.7 (0.4)** | **-5.5 (4.5)** | **-4.4 (3.9)** | **2.4 (4.4)** | **1.4 (3.0)** | **1.1 (5.9)** |

Apart from the above experiments, we have also concerned on risk indicators: volatility, max drawdown, and Sharpe ratio.

### 5.5 Visual analytics

In our study, we further review our findings through a detailed visualization of the predictions generated by our *Hidformer* model for different datasets. As depicted in on sample Figure 4, Figure 5 and Figure 6 the *Hidformer* demonstrates a robust capability to accurately discern different directions of market trends. This ability is of importance for augmenting algorithmic strategies that assist in human investment decision-making processes. It may underscore the *Hidformer* model's potential as a valuable tool for investors seeking to navigate the complexities of the stock market through informed, data-driven strategies.

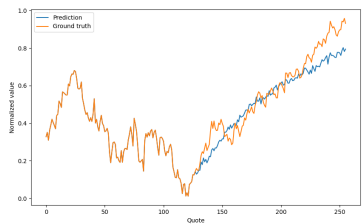

**Fig. 4.** Upward trend prediction.

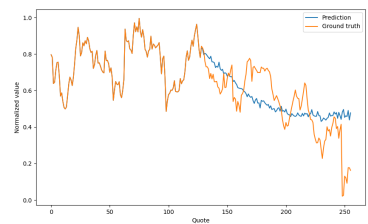

**Fig. 5.** Downward trend prediction.

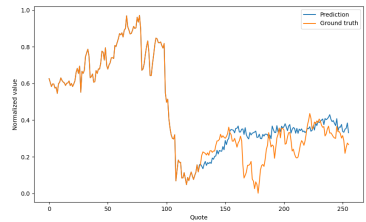

**Fig. 6.** Mixed Trend Prediction: Upward, Sideways, Downward.

## 6 Conclusion and future work

The problem of forecasting stock prices has a particular resonance in the domain of artificial neural networks, where Transformer-based architectures, have been increasingly recognized for their potential in time series prediction. Our research enriches this field of study by empirically validating the capabilities of an extended *Hidformer* model. We have verified the *Hidformer* adaptability and its promising application in financial time series forecasting, highlighting its capacity for enhancing decision-making process in daily trading operations.

Looking ahead, deeper exploration of the *Hidformer* model across diverse datasets and market conditions could yield richer insights into its robustness and efficiency. This would involve using two-dimensional time and frequency towers to process stock prices and volumes separately, with decoders for each sector (or a one-hot vector conditioning one decoder). This approach accommodates daily stock trading specifics, enabling encoders to learn various formations while minimizing the risk of decoders learning contradictory patterns across industries. Comparing the proposed method with the *Informer* model [13, 16] and the reinforcement learning trading strategy [15] would also be worthwhile.